**Title:**
**Pharmacogenomics in the Age of GWAS, Omics Atlases, and PheWAS**

Ari Allyn-Feuer[1], Gerald A. Higgins[1] & Brian D. Athey[1,2,3,4]

1: Department of Computational Medicine and Bioinformatics, University of Michigan Medical School
2: Department of Internal Medicine, University of Michigan Health System
3: Department of Psychiatry, University of Michigan Medical School
4: Michigan Institute for Data Science, University of Michigan Office of Research

*Corresponding author:
Brian D. Athey, Ph.D.
Department of Computational Medicine and Bioinformatics
University of Michigan Medical School, 100 Washtenaw Avenue, Ann Arbor, MI 48109
1-734-615-5774
bleu@med.umich.edu

Funding declaration and acknowledgements:

Mr. Allyn-Feuer was funded by NIH T32 Pre-doctoral training grant 5T32GM070449-12. A.A-F, G.H., and B.A were supported by UM Medical School Departmental Funds. There are no conflicts of interest to disclose. The authors wish to thank Mr. Alexandr Kalinin for helpful discussions.

**Abstract:**

The search for causative pharmacogenomic loci is being transformed by integrative omics pipelines, but their outputs have only begun being applied to test design.  We assess the direction of the field in light of Biobanks/PheWAS, omics atlases, and AI.  We first assess the potential of recent epigenome and spatial genome concepts, datasets, and methods to improve the functionality of PIP-style pipelines.  We then discuss new potential methods of genetic test design on the basis of the outputs of such pipelines.  We conclude with a vision for a pharmacophenomic atlas, in which omics atlas data, PheWAS associations, and biobank data would be used with AI to design thousands of genetic tests for clinical deployment in an automated parallel process.

**Executive Summary:**

**Pharmacoepigenomics: Interpreting GWAS with the Epigenome and Spatial Genome:**

- Pharmacogenomics has failed to create clinically useful genetic tests for many heritable phenotypes, due to problems of scalability, missing heritability, and tractability.
- In other subdisciplines of genomics, these problems are being attacked by interpreting GWAS and PheWAS results with epigenome data from omics atlases to search for causative regulatory variants
- The investigation of pharmacological phenotypes through the "pharmacoepigenome" lens of spatiotemporally regulated gene expression has yielded powerful loci for many phenotypes

**Epigenome Variant Pipelines for Pharmacogenomics:**

- The use of epigenome and spatial genome data to find causative regulatory variants in GWAS significance regions has been accomplished by pipelines including RegulomeDB and HaploReg
- After the demonstration of the value of these methods in pharmacogenomics, the Pharmacoepigenomics Informatics Pipeline (PIP) was introduced
- The PIP and precursor methods have discovered new regulatory variants governing response and adverse events for many psychotropic drugs, as well as anticoagulants

**Future PIP Featureset:**

- A new PIP featureset, still based on the Five Box Model of regulatory variant discovery but incorporating new methods and datasets, can make PIP-style variant discovery much more powerful
- Important new methods include finding target genes with Hi-C data and QTL screens, and using machine learning for tissue specific variant dependence analysis
- Due to limitations in both training datasets and current machine learning methods, the role of machine learning in such a pipeline will of necessity be limited to intermediate roles for the foreseeable future

**Automated Test Design and the Pharmacophenomic Atlas:**

- The dimensionality reduction afforded by the PIP makes cohort-based supervised machine learning on genomic and clinical features a tractable approach for pharmacogenomics test design
- The mechanistic validity of PIP-discovered variants and the clinical validity of a finished classifier will be potent factors in smoothing the way for regulatory approval of such tests, compared to whole-genome machine learning based classifiers
- With PheWAS and biobanked EMRs, the tools are emerging to design thousands of genetic-clinical predictors in parallel, and deploy such a Pharmacophenome Atlas in clinical decision support systems

**The Unfulfilled Promise of Pharmacogenomics**

A common sentiment among clinicians is that Pharmacogenomics, while conceptually appealing, has not lived up to its promise and remains limited in clinical application. While Pharmacogenomics scientists may dispute the pessimistic tone of much comment on Pharmacogenomics, preferring instead to emphasize the many important loci the field has discovered, as well as its noteworthy but isolated clinical successes, it is undisputable that there exist many heritable phenotypes of clinical importance for which no genetic tests are available or for which available tests are not widely used.

This is so primarily for three reasons, all relating to the current methods of Pharmacogenomic locus discovery and test design and validation: scalability, missing heritability, and tractability. Firstly, there are many pharmacological phenotypes which have not been studied with the cumbersome system-specific and locus-specific tools of traditional Pharmacogenomics, e.g. single system GWAS, mechanistic model system experiments, and prospective clinical trials. Secondly, systems which have been studied frequently exhibit "missing heritability," the divergence between top-down (genetic) and bottom-up (genomic) heritability estimates, and consequently discovered loci cannot sufficiently predict these phenotypes despite their heritability. Finally, even for systems wherein important loci have been discovered, the path of translating loci into tests, and tests into the clinic, is lengthy and expensive, and frequently goes untrod.

Although the discipline of Pharmacogenomics has sometimes operated at a conceptual and professional remove from the rest of genomics, related disciplines have faced similar challenges and developed potent tools to address them, many of which have been underapplied in Pharmacogenomics. The application of integrative multiple omics analysis to Pharmacogenomics under the rubric of the Pharmacoepigenome has made a significant impact on Pharmacogenomics, including new locus and network discoveries for many phenotypes, and new tools for discovering them, as well as nascent predictive models. But in the coming decade, the powerful new capabilities enabled by emerging resources from other areas of biomedical research will do even more to unlock the latent potential of the genome in personalizing medical decisions. These technologies include biobanks which fuse biosamples (and, increasingly, omics data) with longitudinal electronic medical records, Phenome Wide Association Studies (PheWAS), deeper atlases of biomedical omics data, and rapidly advancing capabilities in artificial intelligence.

This perspective is divided into three parts. The first covers the broader context of the use of the epigenome and spatial genome to interpret GWAS, and the ways that this knowledge has been translated to Pharmacogenomics with omics pipelines like the Pharmacoepigenomics Informatics Pipeline (PIP). The second covers a future vision for a next-generation PIP-style pipeline which is designed to benefit from the latest discoveries about the spatial genome and epigenome, and from current and upcoming datasets. And the third and final section describes a future vision and methods for the use of mature PIP-style pipelines to design genetic tests using AI methods on cohort datasets, and eventually to do so in parallel with the outputs of PheWAS, enabling genetic prediction of a broad array of biomedical phenotypes to make itself a routine presence in the clinic.

**The GWAS Interpretation Challenge**

Genome wide association studies (GWAS) **[Visscher et al 2017]** have become a cornerstone technique of biomedical locus discovery over the last twelve years, and all signs point to the continuing escalation of this trend. The parallel measurement of millions of SNPs throughout the genome on a microarray at tractable cost **[LaFramboise 2009]** has allowed the traditional methods of genotypic association studies to be carried out in parallel across the entire genome. The resulting explosion in locus discovery for many systems has yielded fundamental discoveries in every area of biology and medicine **[Visscher et al 2017]**. As a result of this, the GWAS catalog **[MacArthur et al 2017]** has swelled to contain over five thousand GWAS from the published literature, while industrial concerns have amassed large numbers of proprietary GWAS.

This trend shows no sign of stopping and every sign of accelerating. Over the last five years, the genetic association methods of GWAS have been applied to two additional forms of high throughput biomedical data analysis, which add additional dimensions of parallelization across many phenotypes. In the case of molecular quantitative trait locus (mQTL) screening **[Delaneau et al 2017]**, GWAS is performed in parallel for a collection of molecular QTLs spanning the entire genome. This may include, for example, gene expression QTL analysis **[Gilad et al 2008, Peters et al 2016]**, but also including an ever-expanding array of molecular phenotypes like DNA methylation **[Banovich et al 2014, Hannon et al 2016]**, DNase accessibility **[Degner et al 2012]**, histone acetylation **[McVicker et al 2013]**, etc. And in the case of Phenome-wide association study (PheWAS) **[Denny et al 2010]**, GWAS methods are parallelized across a large number of medical phenotypes (disease diagnoses, drug responses, physiological measurements, etc) extracted at scale from an electronic medical record **[Hebbring et al 2015]**. The amount of GWAS data available, the array of phenotypes tested, the amount of undiscovered insight latent in these experiments, and the interest in interpreting them will continue to grow.

At the same time, however, the results of GWAS have not lived up to some of the original excitement. At the inception of the technique after the sequencing of the human genome, it was believed by many that the GWAS technique would lead to the recovery of the bulk of the genetic heritability of studied phenotypes **[Collins et al 2001, Ganguly et al 2001]**. As the GWAS techniques matured, however, it became clear that for most complex phenotypes, discovered loci accounted for only a fraction, often a minority, of the heritability of the phenotype **[Slatkin et al 2009, Manolio et al 2009, Gusev et al 2013]**. And in some cases, well powered GWAS recovered few or no significant loci for a heritable trait **[MacArthur et al 2017]**. This "missing heritability" problem has been the topic of perennial debate, with opinion settling around a number of hypotheses:

1) That the missing heritability is accounted for largely by independent causal variants with very small effect sizes below the detection threshold of even powerful GWAS, known as the "gold dust" **[Shi et al 2011]** or "omnigenic" hypothesis **[Boyle et al 2017]**.

2) That the missing heritability is accounted for largely by rare variants not covered in individual GWAS, known as the rare variant hypothesis **[Schork 2009, Gibson 2011]**.

3) That the missing heritability is accounted for largely by cooperative epistatic interaction between multiple loci, rather than linear summation of the statistical predictiveness of individual loci, known as the epistatic hypothesis **[Zuk et al 2011, Sivakumaran et al 2011]**.

4) That the missing heritability is accounted for largely by direct epigenetic inheritance not mediated by DNA sequence, known as the epigenetic inheritance hypothesis **[Slatkin 2009, Franklin et al 2010]**.

5) That the missing heritability is largely illusory, the result of shared environmental factors between family members creating concordant phenotypic outcomes which inflate heritability estimates, known as the environmental hypothesis **[Gage et al 2016]**.

Despite the enduring popularity of the environmental hypothesis in the lay press **[Rossiter 1996, Feldman et al 2018]**, detailed investigation of family members of different levels of relatedness, raised separately or together, from the same and different pregnancies, have decisively refuted this hypothesis for a number of well-studied phenotypes **[Plomin et al 2015, Plomin et al 2018]**. Less clarity has emerged on the topic of the omnogenic, rare-variant, epistatic, and epigenetic hypotheses. It is likely that each of these factors contributes to the heritability problem to some extent and that such extents vary for different types of phenotypes, but this landscape remains murky.

Interpreting GWAS results is a very difficult problem. It involves a number of major challenges, principally discerning the reality of associations, attributing causal character to them, finding the causal variants within linkage regions, and discerning their function.

**The Epigenome**

During the same period of time, array- and sequencing-based assays for a large number of epigenome features, and experiments and atlases conducted with such assays, have revealed that the multifaceted, tissue-specific epigenome has a relationship with gene expression, cell fate, organismal function and dysfunction, and disease, which is both intricate and powerful.

In contrast to the genome, which is relatively well defined, the epigenome comprises a large and growing number of modalities measurable with various assays, most of which take the form of "tracks" comprising numeric levels of observed signal at various positions in the genome, and often a genome-wide vector of signal. These include gene expression (RNA microarrays and RNA-seq, and proteomics) **[Lonsdale et al 2013]**, DNA methylation (assayable with MeDIP-seq **[Staunstrup et al 2016]**, RRBS **[Yong et al 2016]**, WGBS **[Olova et al 2018]**, and other methods), hydroxymethylation (hydroxylmethylation WGBS) **[Huang et al 2010, Wen at al 2016]**, chromatin accessibility (DNase-seq **[Song et al 2010]**, ATAC-seq **[Buenrostro et al 2015]**), histone post-translational modifications principally including acetylation and mono-, di-, and trimethylation of lysine residues at H3K27, H3K9, H3K36, and H3K4 **[Roadmap Epigenomics Consortium 2015]**, but also including a large and growing number of subsidiary histone marks (ChIP-seq, ChIP-Exo), transcription factor binding for hundreds of transcription factors (ChIP-seq), and others. In addition to this are epigenome assays relating to the spatial and functional genome which express themselves as contacts and relationships in squared and other higher-dimensional genome spaces, including molecular QTL (mQTL) screening, enhancer mapping, and spatial genome measurements including 3C **[Dekker et al 2002]**, 4C **[Simonis et al 2006]**, 5C **[Dostie et al 2006]**, Hi-C **[Lieberman-Aiden et al 2009]**, ChIA-PET **[Li et al 2014]**, Genome Architecture Mapping **[Beagrie et al 2017]**, Hi-ChIP **[Mumbach et al 2016]**, and SPRITE **[Quinodoz et al 2018]**.

Epigenome elements differ from cell type to cell type and across the cell cycle in multicellular organisms **[Roadmap Epigenomics Consortium 2015]**, and from person to person **[Flanagan et al 2006, Schneider et al 2010]**, and according to physiological, environmental, and medical conditions **[Schneider et al 2010]**. Thus, the space of possible assays dwarfs any real dataset. Nevertheless, there have been increasingly systematic attempts to produce comprehensive spanning sets of epigenome data with standardized methods, yielding a set of epigenome "atlases" under the rubrics of ENCODE **[ENCODE Project Consortium 2012]**, the Epigenome Roadmap **[Roadmap Epigenomics Consortium 2015]**, the International Human Epigenome Consortium **[Stunnenberg et al 2016]**, and the upcoming Human Cell Atlas **[Regev et al 2017]**, as well as more focused efforts from many quarters. IHEC data now includes a set of core epigenome marks for over a hundred tissues throughout the human body. As a result of this, despite the inherent sparsity of any real dataset, the epigenome is increasingly regarded like the reference genome: as a resource to be consulted for systems and loci of interest, rather than an unknown quantity to be queried experimentally in specific contexts.

The epigenome atlases have identified a set of "core" epigenome elements which determine a set of chromatin states corresponding to the various categories of regulatory states (for genes) and regulatory elements (for noncoding regions of the genome). The most influential method for calling chromatin states is ChromHMM **[Ernst et al 2012]**, which uses a hidden Markov model on "core" epigenome tracks to call fifteen chromatin states including seven types of promoters/enhancers and eight types of activity/repression states. Chromatin states which emerge in particular genomic locations in particular tissues have become a widely used and powerful guide to the functions of the host loci and the tissues in which they operate.

In addition to this, the validation of enhancer and promoter elements with reporter assays based on the transcription of their target genes under the influence of element excision with CRISPR genome editing **[Lopes et al 2016, Gasperini et al 2017, Klein et al 2018]**, what is referred to as "CRISPR validation" of an element, has been increasingly used. Under the influence of these methods, researchers are increasingly able to assess the function, activity, and targets of regulatory elements in a tissue-specific manner by consulting these resources, even in the absence of any purpose-specific experiments.

In addition to this, a powerful set of tools have emerged that use machine learning and explicit algorithms to make predictions about the effects of sequence changes and other perturbations on epigenome outcomes. These have included both explicit algorithms and machine learning methods for predicting TF binding, machine learning methods for predicting variant effects on epigenome tracks and chromatin states, and machine learning models for predicting enhancer targets.

Position weight matrices (PWMs) **[Stormo et al 1982]** have been used to define transcription factor motifs since the 1980s. Despite their algorithmic crudity, their performance at defining binding sites has been surprisingly strong relative to competing methods, and they have the advantage of being very simple to define and use, and very interpretable. With the sequencing of the genome and genome wide ChIP-seq experiments for many transcription factors, it became possible to define motifs for large collections of TFs **[Liefooghie et al 2006]** and assay them genome wide. With the reference and alternate alleles of SNPs, it is thus possible to gauge the comparative adherence of the flanking sequences to the TF under the influence of the various alleles of the SNP. This is performed at scale by algorithms like TFM-Scan **[Liefooghie et al 2006]** and MotifDB **[Shannon et al 2018]**. In addition to this, recent work by Nishizaki **[Nishizaki et al 2017]** and others has focused on predicting the occupancy of TF binding sites by reference to both variants and the chromatin states exhibited by the host loci in tissue-specific data.

In addition to this a large number of machine learning algorithms have been published and widely used that predict the influence of variants on epigenome tracks, especially chromatin accessibility, and on chromatin state. Support vector machines based on gapped k-mers and trained on positive and negative sequences from a specific epigenome track emerged as a potent predictor of variant effects on chromatin accessibility beginning with gkm-SVM **[Gandhi et al 2016]**, deltaSVM **[Lee et al 2015]**, and lsGKM **[Lee 2016]**. But in addition to this, deep neural networks including DeepSEA **[Zhou et al 2015]** and Basset **[Kelley et al 2016]** have been trained on collections of epigenome tracks in a large number of tissues to predict variant effects on epigenome tracks and chromatin states in specific tissues, and have exhibited performance increases over SVM-based methods, including the ability to attain increased speed of training by transfer learning from a trained multi-tissue model to a single-tissue model.

Finally, integrative computational methods for predicting enhancer target genes on the basis of tissue-specific integrative epigenomics and Hi-C data have proved extremely effective at predicting proximal enhancer targets **[Spurrell et al 2016, Hardison et al 2014]**. Such methods have prominently included TargetFinder **[Whalen et al 2016]**, which took an integrative approach, as well as methods of Lieberman-Aiden et al **[Durand et al 2016]**, which are based purely on Hi-C and sequence data.

**The Spatial Genome**

The period since the sequencing of the genome has also seen two epochal discoveries advancing in parallel: firstly that the spatial organization of the genome inside the 3D space of the nucleus varies from cell type to cell type **[Rao et al 2014]**, cell to cell **[Nagano et al 2013, Stevens et al 2017, Ramani et al 2017]**, condition to condition **[Chen et al 2017]**, and over time **[Seaman et al 2018]**, and has a powerful and intricate connection with gene expression, cell fate, and organismal function and dysfunction including disease, and secondly that such organization can be measured in parallel across the multidimensional vector spaces over genomic position which represent combinatorial genomic interactions **[Quinodoz et al 2018]**.

Prior to the emergence of genome sequencing based biochemical assays, the existence of a role for genomic spatial organization was already known from imaging studies and the biochemical methods of chromosome conformation capture (3C) **[Cremer et al 2000, Dekker et al 2002]**, to some extent. It was known that chromosomes in the interphase nucleus segregated into chromosome territories (CTs) resolvable with fluorescent in-situ hybridization (FISH) probes **[Cremer et al 2000]**, that the loci involved in common translocations in cancer were often spatially close to each other in normal interphase nuclei of the tissues from which the cancers arose **[Lee et al 1993]**, and that transcriptional activity demarcated by RNA polymerase II (Pol2) took on a punctate form inside the nucleus **[Hughes et al 1995, Eskiw et al 2008]**.

But such awareness took on a new height with the advent of new measurement methods. Techniques for labeling, imaging, and interpretation all took on new power. Not only did FISH probes of the cell nucleus take on a new precision with the availability of genome-wide collections of bacterial artificial chromosome (BAC) libraries for the easy generation of FISH probes for arbitrary loci **[Baumgartner et al 2006]**, but oligomer FISH (Oligo-FISH) methods **[Schmitt et al 2010]** became easier and cheaper to perform with the availability of the reference genome, bioinformatics tools for probe design, and the rapidly plunging cost of de novo DNA synthesis. The wider availability of 3D confocal microscopes and more powerful imaging methods like 3D SIM **[Shao et al 2011]**, PALM **[Betzig et al 2006]** and STORM **[Rust et al 2006]** super resolution microscopy, labeling methods for multiple probes at the same time, spectral deconvolution **[Zimmerman et al 2003]** and white light lasers **[Chiu et al 2012]** to make channels easier to separate, and other methods made it possible to discern label positions better than ever before. CRISPR imaging **[Chen et al 2013, Chen et al 2014]** made it possible to image nuclear labels in live cells. And the proliferation of 2D and 3D image parsing packages, including the work of Rajapakse et al **[Rajapakse et al 2011, Seaman et al 2015]** and Misteli et al **[Shachar et al 2015, Jowhar et al 2018]**, as well as Kalinin et al **[Kalinin et al 2017]**, made it easier to extract high content semantic information from nuclear architecture imaging.

This imaging work confirmed the earlier explorations of the pre-genome era and added new discoveries. It was discovered that the punctate elements of Pol2 activity were in fact collections of co-regulated genes from different chromosomes being transcribed and regulated together **[Papantonis et al 2013]**, that they assemble stochastically on timescales of minutes for transcriptional bursting **[Ghamari et al 2015]**, that they are located on the periphery between chromosome territories and preferentially located in the center of the nucleus **[Cremer et al 2015]**, and that their assembly predates transcription **[Krijger et al 2017]**. It was discovered that repressive lamin-associated domain (LAD) elements in the genome are transcriptionally repressed **[van Steensel et al 2017]** and that escape from transcriptional silencing depends on spatial escape from the lamin **[Robson et al 2017]**. It was discovered that proximal promoters and controlling enhancers unite spatially for transcription **[Rao et al 2014]**, including super enhancers controlling collections of co-regulated genes **[Thibodeau et al 2017, Gong et al 2018, Huang et al 2018]**. And experiments in serial FISH of a collection of FISH probes along the length of a chromosome showed that the spatial compaction of collections of loci within a collection of loci recapitulates their contact frequencies as measured by 3C-based methods **[Wang et al 2016]**.

And the developments in sequencing-based methods for gauging chromatin spatial information were even more profound. The sequential development of parallel 3C-based methods including 4C (one genomic location against the genome) and 5C (a collection of probes against each other) reached a disjunction with the development of Hi-C: high throughput chromatin conformation capture. It involves cross-linking fixed cells and digesting the genome with a restriction enzyme (or, for Micro-C, mainly used in small genomes, an unselective endonuclease like MNase), followed by religation, sonication, and paired end sequencing. The result is a paired end sequencing library wherein the genomic locations of the paired reads do not correspond to elements close to each other in sequence space, but in physical space. Such reads can be compiled into a chromatin contact map, potentially of the entire squared genome.

This method, which allowed parallel 3C-style measurement of the entire genome against the entire genome, producing maps of chromatin contacts in squared genome space, electrified the field of chromatin structure and became the standard in genome structure research. Progressive protocol optimization has allowed the methods to produce higher quality data with lower effort, and to address a wider collection of cell lines and tissues. Experiments in a large number of cell lines and conditions have highlighted the commonalities in genome organization and the ways that organization differs over time. Hi-C data is increasingly being used for fundamental genomics tasks like assembling reference genomes **[Korbel et al 2013]**, phasing haplotypes **[Ben-Elazar et al 2016, Selvaraj et al 2013]**, and detecting chromosome translocations **[Chakraborty et al 2017]**.

But even more than this, Hi-C data has been produced a revolution in our understanding of spatial genome organization, the largest component of which has been the discovery of topologically associating domains (TADs) **[Dixon et al 2012]** in the human genome and other genomes. These self-associating regions have been identified as potent spatial and functional elements in the human genome. The division of approximately 80% of the human genome into approximately 2500 TADs is remarkably robust, being largely conserved between cell types in the human body **[Rao et al 2014]**, between different humans **[Ruiz-Velasco et al 2017]**, and under disease states **[Rao et al 2014]**. In fact, syntenic regions of related genomes (e.g. mouse) often share the same TAD structure as the related regions of the human genome **[Krefting et al 2017, Nora et al 2013]**. TADs also function as replication domains **[Pope et al 2014]**. Moreover, TADs mediate long range spatial interactions **[Rao et al 2014]**: the contact frequency in any given portion of the squared genome will more closely correlate with a more sequence-distant portion which is in the same TAD pair than a sequence-proximal portion spanning TAD boundaries.

While the portion of a genome which composes a TAD is relatively invariant, TADs differ from one cell type and biological condition to another in their degree of transcriptional activity. This differentiation between the "A" and "B" compartments is connected with the sign of the dominant eigenvector of a genome-wide Hi-C matrix **[Dekker et al 2013]**, the degree of spatial openness as observed by both Hi-C, imaging, and biochemical experiments **[Roadmap Epigenomics Consortium 2015]**, the degree of gene expression **[Roadmap Epigenomics Consortium 2015]**, the presence of active histone marks **[Roadmap Epigenomics Consortium 2015]**, the presence of activating transcription factors **[Roadmap Epigenomics Consortium 2015]**, replication timing **[Pope et al 2014]**, and the extent of long range and interchromosomal contacts **[Rao et al 2014]**. Genes located in the same TAD tend to be co-regulated, and they are regulated partially by their TAD context.

Moreover, the sequence context governing these regulatory interactions is beginning to illuminate under sustained investigation. High resolution Hi-C experiments have discovered a hierarchy of

super- and sub-TADs running all the way down to the level of "loop domains" **[Rao et al 2014]** which spatially unite the proximal promoters of genes with their intra-TAD proximal enhancers. These contacts, and those above them in the hierarchy, are largely governed by the presence of convergent pairs of CTCF sites **[Nichols et al 2015, de Wit et al 2015]** which form a CTCF-cohesin anchor binding loci together by means of a loop extrusion mechanism which has been verified by biochemical and imaging methods **[Fudenberg et al 2016, Sanborn et al 2016]**. Perturbations of these sequence elements by CRISPR in vitro, or by disease-related mutations in vivo, as for example in enhancer hijacking in cancer, have the predictable effects on Hi-C maps, the epigenome, and gene expression **[Wutz et al 2016, Sanborn et al 2016]**.

The march of Hi-C in genomics shows every sign of continuing to escalate, with the number and variety of Hi-C experiments escalating year by year. Recent work shows that genome assembly benefits from Hi-C data **[Korbel et al 2013]**, so that in the future, Hi-C library preparation may be used for routine genome sequencing, making Hi-C data available in larger numbers of samples than any other epigenome modality (aside from gene expression). In addition, the upcoming Human Cell Atlas **[Regev et al 2017]** is widely expected to feature in-depth Hi-C data on every major cell type in the human body. Single cell Hi-C is shining a light on the variation of cellular organization between cell types and over the cell cycle. Promoter capture Hi-C **[Mifsud et al 2015]** is being explored as a medical diagnostic **[Mishra et al 2017, Baxter et al 2018]**.

And new methods of gauging chromatin organization with sequencing continue to proliferate, offering new discernment of various features. Micro-C **[Hsieh et al 2015]**, using endonucleases instead of restriction enzymes, has carried the resolution of Hi-C all the way down to the nucleosome level in small genomes (e.g. yeast and drosophila). SPRITE **[Quinodoz et al 2017]**, which uses serial dilution and combinatorial labeling to uniquely label suspended chromatin complexes before single ended sequencing, is producing deeper Hi-C style maps with less sequencing, and also allowing the exploration of multiple-locus contacts in higher order combinatorial genome spaces. Genome Architecture Mapping **[Beagrie et al 2017]**, which uses single cell sequencing of cryosectioned nuclei followed by statistical modeling of the co-occurrence of pairs of sequence regions, cannot produce high resolution intra-TAD information, but generates Hi-C style contact maps with less sequencing, and also can be used to produce calibrated physical distances.

This proliferation of data is increasingly being used to form three dimensional models of chromosome territories and even the entire nucleus. Early work in this area met with difficulty due to low resolution data, the complexity of the task and the overconstrained nature of the data, and the challenges caused by structural heterogeneity **[Nagano et al 2013]**. This may be clearly seen by the preeminence and then rejection of the "superaxis" model of chromosome territory structure, in which the sequence of a chromosome is approximately recapitulated by the spatial ordering of TADs along a "superaxis," with the string of tads threading back and forth between the A and B compartments of the territory, possibly by means of a coil. Multiple spatial modeling investigations based on different modeling methods and different Hi-C datasets independently discovered the superaxis **[Hu et al 2013, Nagano et al 2013]**, but it was subsequently demolished by the publication of serial FISH data on threading TAD positions in chromosome territories in single cells **[Wang et al 2016]**. The latest modeling methods use single cell data to refine models based on higher-resolution ensemble Hi-C **[Lando et al 2018]**, and do not produce superaxis behavior in the models. They show better accord with 3D FISH data than prior models.

**Integrative Epigenome Models of Variant Function Applied to Association Hits in Pharmacogenomics**

The combined power of multiple epigenome modalities to interpret variant function began being exploited to address the interpretation challenges of GWAS around the time of the publication of the Phase 2 ENCODE maps **[ENCODE Project Consortium 2012]**, and was in wide development by the time the Epigenome Roadmap **[Roadmap Epigenomics Consortium 2014]** data landed. The most visible sign of this was in the variant annotation and interpretation methods being published, prominently including the early contributions of RegulomeDB **[Boyle et al 2012]** and HaploReg **[Ward et al 2012]**.

These pipelines work by annotating SNPs multimodally with information from multiple types of omics information. In the case of RegulomeDB, this included DNase sensitive regions, validated promoter and enhancer regions, transcription factor binding sites, and predicted regulatory elements. HaploReg expanded on this suite by adding tools to gauge the relevance of SNPs in the area of a lead SNP with linkage, and looking at PWM perturbation by SNPs, deeper epigenome data from the Roadmap, and eQTL data. Both have been widely used.

All such models rely on an overall paradigm of genomic regulation sometimes described as the "pharmacoepigenome," **[Higgins et al 2015]** in which, by virtue of the facts that 1) associations arise from causative regulatory variants influencing the underlying genomic machinery of a phenotype, 2) regulatory variants and their targets can be identified and parsed by looking for the hallmarks of regulation in epigenome datasets, and 3) the machinery underlying genetic variation in a phenotype is often joint with the machinery underlying the phenotype itself and related phenotypes, it is concluded that powerful insights into the mechanisms of a wide variety of phenotype, not necessarily constrained to pharmacogenomics, can be obtained by looking for tissue-specific regulatory variants and their targets in the regions surrounding association hits for a collection of related phenotypes.

Such models have been used to interpret GWAS, looking in significance regions for regulatory variants, by a large number of different methods mostly comprising individual ad-hoc analyses. They differ significantly but largely adhere (fully or partially) to a number of common themes:

1) Analyzing multiple GWAS on the same (or related) phenotypes together in the same analysis.

2) Analyzing many variants, not just lead SNPs. For primary GWAS this is often done with the significance region or with a p-value fold change cutoff from the lead SNP. In secondary analysis, it is often done by linkage. Some analyses have used sequence distance cutoffs, but this is not biologically informed and is inadvisable.

3) Looking for regulatory variants with tissue-specific data that is relevant to the phenotype under investigation.

4) Attempting to identify the target genes of candidate regulatory variants discovered in the analysis.

5) Filtering, ranking, and organizing the result genes together with pathways and ontologies, looking either for genes of preexisting known relationships to the phenotype, or genes that cohere with each other, e.g. by shared pathway or ontology membership, or coregulation by a known transcription factor.

Such analyses have become more common, with epigenome-based causal variant methods appearing in even primary GWAS analyses.

**The Paucity of Pharmacoepigenomics in Clinical Practice**

By contrast, the discipline of pharmacogenomics has historically made little use of any of the advanced variant discovery and interpretation methods discussed above. To date, the most widely used biomarkers for clinical pharmacogenomic testing are a set of pharmacokinetic (PK) gene variants located in CYP genes, encoding the main drug metabolizing enzymes, while only a small family of pharmacodynamic (PD) genes have been utilized, and regulatory variants, even less **[Higgins et al 2015]**. While such tests often achieve clinical utility and cost-of-care reduction **[Higgins et al 2015]**, they often account for a minority of the inter-individual genotypic variation in important drug-related phenotypes **[Higgins et al 2015]**. By and large, however, these tests are designed manually using variants in a small pool of candidate genes.

Although pharmacogenomic phenotypes have been investigated with GWAS since its inception **[Giacomini et al 2017]**, most available pharmacogenomics tests continue to be based on highly penetrant coding variants revealed by gene-specific work, with GWAS findings, PD genes, and regulatory variants persistently underutilized. Deployment of GWAS in pharmacogenomics variant discovery has lagged deployment in other disciplines, epigenomic interpretation of GWAS results has been underutilized, and the translation of GWAS results into clinical tests has been slower still in most areas.

Indeed, much pharmacogenomics variant discovery still proceeds along traditional lines involving the search for coding variants to be designated as star (*) alleles, with a particular emphasis on PK genes for absorption, distribution, metabolism, and excretion (ADME). Such genes have formed the focus for test development for response, dosing, and adverse drug events (ADEs) and adverse drug reactions (ADRs).

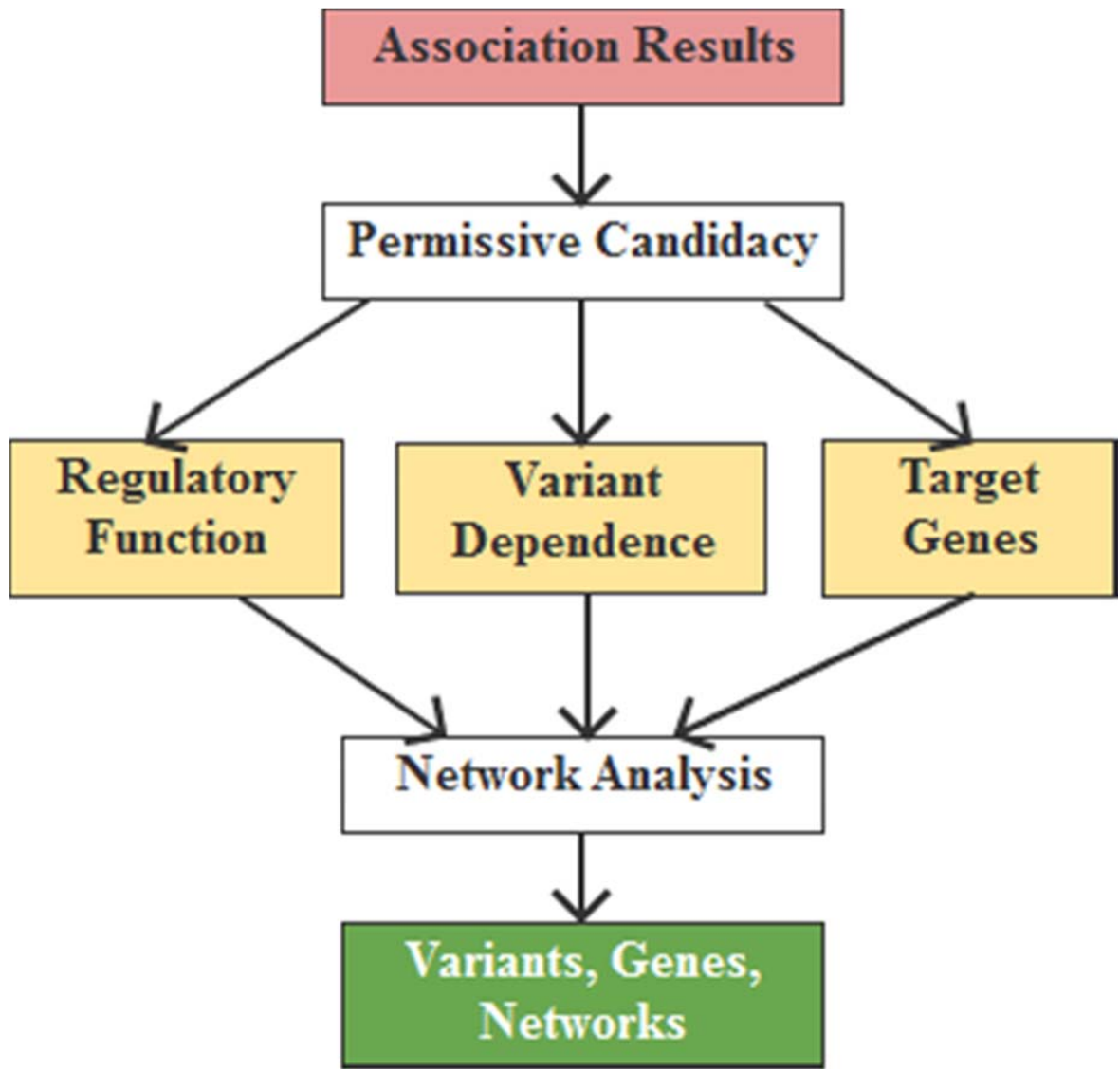

**Figure 1: The Five Box Model**

**The Five Box Model and the PIP**

Among the omics pipelines used for variant discovery in GWAS significance regions is the Pharmacoepigenomics Informatics Pipeline (PIP) **[Allyn-Feuer et al 2018]**. The PIP is an integrative multiple omics variant discovery pipeline specifically designed for the reanalysis of GWAS for pharmacogenomics. The PIP's function is conceptualized with the five box model of regulatory variant discovery (**Figure 1**), reflecting a conceptual scheme of five general properties a variant may exhibit which provide evidence, individually and collectively, that it may be a causal regulatory variant in a phenotype. They are permissive candidacy, regulatory function, variant dependence, target genes, and network analysis.

By permissive candidacy, we mean that the variant has in some way come to the attention of a genome-wide screen: that it is located in a population-specific linkage disequilibrium with a variant associated with a phenotype of interest, or that it regulates a gene whose mechanistic importance to the phenotype of interest has already been established.

By regulatory function, we mean that the variant is resident in a portion of the genome which is a regulatory element, a promoter or enhancer, in one or more of the particular tissues which are relevant to the drug-disease system.

By variant dependence, we mean that the function of this regulatory element must be dependent on the status of that variant, through the alteration of sequence features which help to determine the epigenome. Such sequence features may be a specific binding site for a transcription factor, but may also be a more general propensity score for an epigenome feature, as determined by an appropriate bioinformatics algorithm such as a learning machine.

By target genes, we mean the variant must have identifiable target genes with which it is spatially and/or functionally associated, putatively whose expression it regulates.

And by pathway analysis, we mean that we may say of a collection of putative genes and variants for a phenotype or phenotype cluster, that taken as a totality they are associated with each other and with ontologic and/or network categories which are connected to the phenotype under investigation.

These five concepts may be evaluated in different ways under different circumstances: with different datasets, with different algorithms, manually and under automation. Nevertheless they are conceptually durable, and all the pre-PIP workflows, each extant version of the PIP, and our future plans for a PIP-successor pipeline, fall within this overall orienting framework.

**Pharmacogenomics in the Age of GWAS, Omics Atlases, and PheWAS**

The Pharmacoepigenomics Informatics Pipeline and associated tools and methods have added value in the identification of causative enhancer variants for phenotypes of interest, and the identification of their target genes for future mechanistic work. The variants, genes, and networks discovered for warfarin **[Allyn-Feuer et al 2018]**, lithium **[Higgins et al 2015, Allyn-Feuer et al 2018]**, valproic acid **[Higgins et al 2017]**, and other drug-disease systems **[Higgins et al 2015]** can be used directly in traditional test design, and other phenotypes can be investigated with these tools.

However, this featureset is not the last word. Advances in a number of areas, including the underlying biology of enhancer function, artificial intelligence, omics atlases, biomedical ontologies, and genotyped medical records from biobanks and clinical trials have opened the doorway to much more powerful PIP-style pipelines, which are more sensitive to interactions which cannot be detected by the current PIP, and more thorough in pruning out interactions which are not as promising. More, they have opened the door to a future wherein such pipelines are used pervasively in parallel against thousands of biomedical phenotypes, with results that can be used in routine medical practice.

The second portion of this perspective describes an evolved PIP featureset which is based on current cutting edge methods in all the subdomains of biomedical inference from which the PIP draws, and some notions of how to score variants and genes and pathways in an evolved PIP with machine learning. It describes an orienting framework for how to use the output of PIP-style pipelines in biomedical genetic test design. And finally, it concludes with a vision for the use of such a pipeline with biobank datasets and biomedical ontologies to create a genome-wide, phenome-wide pharmacophenomic atlas of predictive models for thousands of phenotypes, which could be then be implemented directly in a clinical decision support system.

The availability of genetic prediction on a routine basis for medically important phenotypes, shortening the translation cycle on genetic discovery, can improve patient care.

**An Evolved PIP Featureset**

The underlying scientific domains of epigenome regulation, spatial genomics, and population genetics on which the PIP is based have not stood still since the PIP featureset was laid down in 2016. They have continued to advance. And while a future evolved PIP featureset will still be based on the overall Five Box Model of regulatory variant discovery, every portion will be affected by these significant discoveries. In addition to this, the universe of data from which a future PIP-style pipeline can draw has expanded vastly.

This will begin with data input: unlike the current PIP, which treats all variant inputs and all tissues uniformly, an evolved PIP would separately take tissue inputs for a collection of related phenotypes, and separate inputs of the relevant variants and populations for each phenotype, as well as information about the degree of relatedness of phenotypes to each other. In addition, the directionality of the variant effect is important: an evolved PIP would track which of the alleles of the SNP had which effect on the phenotype, and in the context of the variant dependence portion of the ERV workflow, the directionality of effects on TF binding and enhancer function will be gauged as well. The concordance of this information within the scores for each SNP, and between the common regulatory SNPs for a gene, will function as important information in the context of scoring.

Such a PIP would have the opportunity to draw on a wealth of data sources on which the current PIP does not draw. In addition to the resources of the current PIP, new and not-previously-available datasets which would add significant value would include:

- Primary GWAS and genotyped cohorts for relevant phenotypes
- Expanded collections of genomes for linkage analysis under the auspices of various biobanks and genomics initiatives
- Expanded omics atlases with deeper omics on more tissues and cell types, under the auspices of the International Human Epigenome Consortium (IHEC) **[Stunnenberg et al 2016]** and the upcoming Human Cell Atlas **[Regev et al 2017]**
- Expanded atlases of TFBS motifs and binding data, under the auspices of MotifDB **[Shannon et al 2018]** and the omics atlases
- Collections of Hi-C data from the Hi-C laboratories (currently highly decentralized), and soon in the form of a bodywide atlas from the Human Cell Atlas
- Expanded collections of molecular QTLs (currently highly decentralized) under the auspices of various biobanks, clinical trials, consortia, and individual analyses
- Libraries of validated enhancers
- The updated pathway mapping libraries of IPA **[Kramer et al 2013]** and other pathway mapping methods

A conceptual featureset for such a pipeline is shown in **Figure 2**, and described in more detail below.

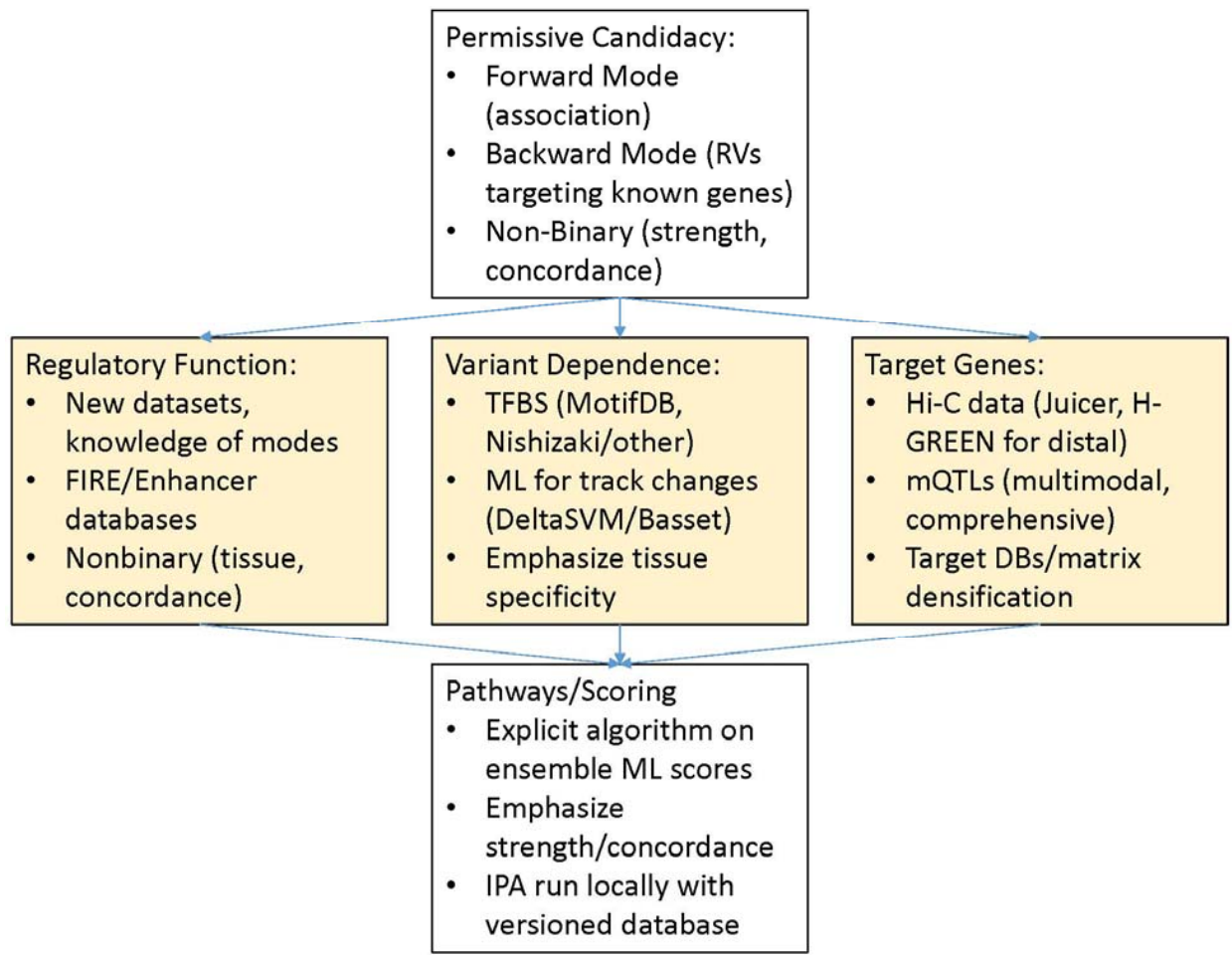

**Figure 2: Conceptual Featureset for a Next-Generation PIP**
Schematic visualization of the envisioned featureset of a next-generation PIP, organized according to the five major elements of the five box model of regulatory variant discovery.

**Permissive Candidacy**

Currently, all PCV identification in the PIP proceeds in a "forward" mode, proceeding from the phenotype toward a locus by association. This is currently done with GWAS lead SNPs by population-specific linkage analysis.

This forward mode of PCV discovery can be expanded. For one, linkage analysis with 1000 Genomes **[1000 Genomes Project Consortium 2015]** populations can be enhanced with the use of more relevant population groups as the number of analyzable genomes continues to expand, and the granularity of population genetics mapping continues to refine. For another, while linkage analysis has been the best approach for secondary analysis of GWAS for which only topline results are available, primary analysis of GWAS and CGAS can offer a more granular approach, when such data are available. Rather than using linkage, the collection of associated variants can be used together, in sum.

Moreover, however, a second mode of PCV identification, the "backward" mode, can take on a more prominent role. In the current PIP, we identify body SNPs of known genes of mechanistic importance in a phenotype as PCVs. In fact, however, what is desired is not body SNPs, but those SNPs which alter the sequence or expression of these genes, i.e. both coding SNPs and regulatory SNPs for this gene, regardless of their position in the genome. Thus, the evolved featureset can take gene inputs, just like the current one, and look for not only coding variants, but also tissue-specific regulatory variants for these genes, no matter where they may be in the genome, using the same target gene tools as the ERV workflow to find PCVs.

Moreover, although the current PIP treats all PCVs equally, PCV status need not be a binary. The strength of a PCV relationship may be modulated by a number of factors. For example, currently all affiliated phenotypes are weighted equally; in the Warfarin analysis, this included not just response and adverse events, but also disease risk and background phenotypes. In an evolved featureset, we would wish to give more weight to variants for more directly related phenotypes.

But for another, the strength and directionality of identification matters. For example, a variant exhibiting a lower p-value in a source GWAS, or stronger linkage to a lead SNP, is probably a stronger candidate. So, too, is a "backward mode" variant with a stronger target gene relationship, or a relationship with a more validated target gene. In addition, a variant identified through both "forward" and "backward" methods may be considered a particularly strong candidate.

**Regulatory Function**

In the current PIP, variants are evaluated for regulatory function in a rigid way: they must have one a promoter or enhancer chromatin state in a relevant tissue, and each of a collection of relevant histone marks in a relevant tissue, and be located in accessible chromatin in a relevant tissue. However, this status is awarded in a simple binary manner, and with no requirement that these tissues be concordant with each other or with other tissue-specific information.

In an evolved featureset, in addition to using larger datasets with higher quality and more relevant data (e.g. on tissues of more granularity), the criterion could enforce concordance between the tissues for each omics modality. It would also serve to look for membership in databases of validated enhancers, and particularly potent categories of FIREs and super enhancers. It would place more emphasis on the features whose relevance has been emphasized in recent work, particularly chromatin accessibility, Hi-C contacts, TF binding, and chromatin state, over individual histone marks and DNA methylation, which have been downplayed.

In addition, of course, this status need not be binary (as in the current PIP), but could scale with the strength of the appearance of regulatory status, the level of inter-tissue concordance, and the degree of relevance of the tissues to the phenotypes for which the PCV attained PCV status.

**Variant Dependence**

The current PIP uses only one modality for variant dependence analysis, and this modality is both limited, and poorly evaluated. This is PWM analysis, using the Kellis et al library of PWMs **[Ward et al 2016]** along with the software TFM-Scan **[Liefooghe et al 2006]** to look for alterations. The rationale for this type of analysis is strong: variants altering TFBS are frequently potent enhancer variants **[Higgins et al 2015]**, and PWMs are a potent means of gauging TF binding affinity **[Stormo et al 1982]**. In addition, the "variants" analysis **[Higgins et al 2015]** showed that concordant changes in TFBS affinity for enhancers located at ChIP-seq-validated binding sites were a good mark of regulatory variant function.

However, the current methods in the PIP are weakened by an obsolete database of PWMs, as well as a crude and unmaintained algorithm for evaluating conformity. In addition, there is a format conversion from probability to frequency matrices which introduces some imprecision. In addition to this, although the current PIP tests for conformity and tests for TF binding, it does not value concordance between these measures.

An evolved PIP featureset would use a comprehensive versioned and updated PWM library in its native format, along with compatible and maintained code for PWM conformity. MotifDB **[Shannon et al 2018]** is one possibility. However, it may also be advisable to consider abandoning PWMs entirely in favor of a machine learning based method for gauging TFBS occupancy as a function of sequence, such as the Nishizaki algorithm **[Nishizaki et al 2017]**, which can gauge both sequence effects and cell type effects (by using epigenome data).

A more significant advance would be the use of machine learning methods to gauge variant effect on both epigenome tracks, chromatin state, and enhancer function. Several machine learning applications have been developed for predicting the impact of non-coding SNPs in GWAS on phenotypes, but fewer than 40% of GWAS publications utilize these tools **[Nishizaki et al 2017]**. There are now many machine learning applications that score features, such as DHS for prioritization of regulatory function and protein annotation of chromatin loops, to predict functional enhancer-promoter interactions and drug-target inference.

Deep learning applications for detection of regulatory elements within the non-coding genome are beginning to emerge **[Angermueller et al 2016, Ching et al 2018, Park et al 2015]**. Since the publication of GKM-SVM **[Ghandi et al 2016]** and its variant DeltaSVM **[Ghandi et al 2014]** in 2014, such applications have proliferated. The use of DeltaSVM to gauge variant effect on chromatin accessibility and therefore enhancer function was used successfully in the valproate variant analysis, and exploration of the PCVs from the warfarin PIP experiment showed that the predictive power of this metric was orthogonal to the PIP. While such applications have proliferated and address a variety of different factors, the type which are most likely to be useful in the context of a PIP-style pipeline would be those directly relating to enhancer function or chromatin state, and which are trained on tissue specific data. This would probably be Basset **[Kelley et al 2016]** or DeepSEA **[Zhou et al 2015]**, or similar methods.

Variants that modify TFBS for TFs with binding activity at those loci, and with predicted allele bias, would have very robust evidence of allele dependence in the regulatory function of the host loci. In addition, concordant directionality between TF binding and enhancer function predictions will be an important form of evidence.

**Target Genes**

The target gene module in the current PIP, which evaluates target genes only by sequence proximity and QTL status (neglecting QTL targets), is inadequate. Both the "variants" and valproate analyses showed the value of finding target genes with Hi-C data, and recent work on target gene analysis with molecular QTLs and Hi-C data, along with machine learning, has been extremely fruitful.

An evolved PIP would locate target genes of regulatory variants with four methods:

1) Databases of CRISPR-validated **[Lopes et al 2016, Gasperini et al 2017, Klein et al 2018]** enhancer targets.

2) A comprehensive genome-wide and tissue-specific collection of molecular QTLs, comprising expression QTLs, DNA methylation QTLs, DNase accessibility QTLs, histone acetylation QTLs, and other modalities of molecular QTLs. Such a library should address major eQTL mapping experiments from GTEx and the national biobanks. A QTL relationship between a SNP and the expression of a gene or the epigenomic status of the gene body is indicative of an enhancer relationship.

3) Hi-C based methods for proximal target identification, such as TargetFinder **[Whalen et al 2016]** and the target finding capabilities of Juicer **[Durand et al 2016]**.

4) H-GREEN **[Allyn-Feuer et al 2018]** or similar methods for distal target identification.

The number, type, and strength of target identifications with these methods will be an important gauge of the strength of a regulatory interaction.

In addition, it may be advisable to consider using the squared genome wide collection of tissue specific enhancer target interactions with matrix densification machine learning methods to densify sparse measures of target genes. Of the methods above, both mQTL mapping and distal Hi-C mapping suffer from significant data sparsity, and densification methods may add value.

**Pathway Mapping**

The pathway mapping functionality in the current PIP relies on the human all-tissue grow and connect functions of Ingenuity Pathway Analysis **[Kramer et al 2013]**, consulted manually on the basis of the output genes. It should be possible to improve on this functionality in several ways:

Firstly, it is important that an evolved PIP's pathway mapping function run locally, run in an automated manner, and use versioned data. IPA licenses are available which offer API access to IPA commands, along with local storage of the database and/or versioned online access to historical quarterly updates. If the evolved PIP is to use IPA or another commercial or open-source pathway mapping tool, such features are essential.

More fundamentally, however, IPA, in evaluating genes only without attention to their origin or their PIP-determined relationships, is in some sense operating in a reductive manner. It can detect whether the collection of genes identified in a PIP experiment have known relationships with the phenotype under investigation or with each other, but it cannot assess the internal relationships among the set of genes and variants as identified by the PIP. In addition to the external, literature and experiment based relationships, it should strike us as important to pathway relationships if, for example:

- The same TFBS is altered and/or the same TF is present at multiple loci for a phenotype
- The same gene is regulated by multiple variants for a phenotype
- The same variant for a phenotype regulated multiple related genes
- The same genes and variants are concordantly identified for a cluster of related phenotypes

While a full exposition of the types of relationships we should seek and evaluate would be essentially a reinvention of the entire field of pathway mapping, and thus outside the scope of this discussion, it will suffice to note that different pathway mapping options may exist, that new functionality recently added to IPA around regulatory variants may be helpful, and that particular needed functionality can be added.

**Scoring**

The scoring algorithm of the current PIP is minimal in nature, comprising an explicit scoring algorithm resulting a binary determination of status for each PCV for each of the components of the five box model, and a simple intersection at the end of scoring to determine a set of intermediate candidates. This method was adopted not because it is optimal but because it is simple.

In the evolved PIP, each component will generate a vector of numeric and qualitative scores of various types, which we wish to integrate into a holistic picture, not necessarily of which variants "pass" a final binary, but of which variants from the set of PCVs are most likely to be predictive, which among them are likely to be mechanistically related, and which genes are likely to play in the mechanism.

Among possible scoring systems, systems in which "points" are awarded for various categoric and numeric elements and summed up to determine a "score" are appealing for their conceptual simplicity, but unlikely to be suited for this complex task, because of the interrelated and cooperative nature of the types of evidence which make up a regulatory variant determination. For example, a variant with no evidence of variant dependence is probably null, regardless of the potency of an enhancer element in which it is located or the target genes of that enhancer. Despite our experiments with "points" based scoring systems, they are unlikely to be fruitful.

The universe of possible explicit scoring algorithms, incorporating "points" and thresholds and sliding scales, is vast. And complex explicit algorithms have achieved great success in many applications in biology, bioinformatics, and medicine. It is possible to envision an ensemble of explicit methods capturing a great deal of complexity and yielding "good" scoring. Nevertheless, it will be useful to consider another possibility: the use of machine learning methods to address this question.

**Machine Learning for PIP scoring: Supervised Classification, Overfitting, and Data Availability**

Most successful deep learning applications rely on large labeled data. While many biological and clinical datasets until recently were limited by amount of available labeled samples compared with the big data analytics applications such as natural image processing and NLP, as the number of samples increases and the number of relevant high-quality labeled datasets expanded, the wealth of pertinent pharmacogenomics data that can be used for analytics is now a big data challenge on par with contemporary applications in other domains. Thus, variant-based learning machines have proliferated and are now widely used in genomics, as described above.

Multimodal, multi-task, and transfer learning are often used to alleviate data limitations to some extent. Transfer learning approaches include training a deep network on a large existing dataset, and then using this pre-initialized model to learn from a smaller dataset, which typically leads to improved performance **[Ching et al 2018]**. When training data is not (fully) labeled, various semi-supervised techniques can be employed **[Ching et al 2018, Iglovikov et al 2017, Beaulieu-Jones et al 2016]**. Data quality is another important concern in deep learning applications. Although deep learning models can be trained directly on raw data, low quality datasets may require additional pre-processing and cleaning. In addition, there is the challenge of preventing overfitting, which requires a very careful design of the model evaluation scheme, including usage of cross-validation techniques, normalization by subjects, and suitable validation metrics.

In the context of the PIP, the consequences of this are clear. There does not currently exist a set of adequately characterized positive controls for pharmacogenomics regulatory variants, of a type, scale, or breadth, to allow the construction of a monolithic supervised machine learning algorithm for scoring PIP variants. Indeed it is not entirely clear from where, with any plausible level of effort, such a training set would come. This complicates matters considerably.

However, subcomponents of the overall model exist wherein training sets for machine learning are available, including the portions described above where machine learning algorithms are used to generate scores in individual modules of the PIP featureset. These include the use of hidden Markov models to generate chromatin states, the use of neural networks to predict variant effects on DNase accessibility and chromatin state, motif discovery by clustering, matrix densification for regulatory variant target discovery, etc. Artificial intelligence algorithms also have been demonstrated for variant imputation.

It may be that databases of validated tissue specific enhancer variants may function as a training set along with omics atlas data, so that the entire ERV workflow (regulatory function, variant dependence, and target genes) may be carried out with a supervised-training machine learning algorithm to predict the presence or absence of a "validated-appearing" enhancer variant.

In addition to this, the relationship between metrics of PCV status, like target gene status, association p-values, etc, and status as a plausible causative variant for the phenotype, among variants which are called as potent regulatory variants, may be addressable with Bayesian statistical approaches constructed synthetically from statistical models in population genetics. Similar approaches may be applied to network membership in the pathway mapping portion of an evolved workflow.

Thus, although it appears intractable to attempt to replace PIP scoring with a monolithic machine learning algorithm using any current tools, it will be possible to gain more insight from more data by the replacement of progressively larger portions of the intermediate scoring (the more defined portions) with machine learning algorithms, even as the overall scoring system remains an ensemble constructed explicitly.

**Using the Output of PIP-Style Pipelines to Develop Genetic Tests**

The output of the PIP is a set of variants, genes, and pathways with a putative causal role in a phenotype of interest. Although such results may be useful for purposes of mechanistic research in the biology of a phenotype, or the search for druggable targets and repurposing opportunities. However, in the context of pharmacogenomics the object of most PIP analyses will be to develop genetic tests (and more broadly, clinical predictive models) for eventual clinical deployment. While an authoritative discussion of this subject is beyond the scope of this thesis, it will serve to discuss this topic briefly in light of the above.

Typically, first generation pharmacogenomics tests were designed by a simple linear combination of variant effect sizes, or with a linear regression model. Later first-generation tests such as the Genesight **[Health Quality Ontario 2017]** psychotropic panel were constructed manually as an explicit combinatorial decision tree algorithm on the basis of effect sizes, paired variant effects, and manual expertise and adjustment. Then, these explicitly constructed tests could be validated against genotyped cohorts with response data before being tested in randomized controlled trials.

The PIP and PIP-style pipelines can be used to design tests in this mode. But doing so would deprive the designer of much of the benefit of the PIP and of modern data sources. With significant pleiotropy and epistasis of pharmacogenomic loci, the reduction of features to a tractable number, and the availability of genotyped cohorts and machine learning methods, it is anticipated that the ideal method for designing a pharmacogenomics test on the basis of PIP results would incorporate these advances.

In this case, a genotyped cohort with genomic information on all the output loci of a PIP experiment, along with clinical variables deemed important in test design, and outcome information, would be gathered, preferably retrospectively from a GWAS or biobank. Then, a machine learning predictor would be trained to predict the outcome variable with this cohort, with cross validation. Since this is a relatively low dimensional space with supervisory information, and because the intention would be to clinically deploy the finished predictor, a relatively simple machine learning method like an SVM or Random Forest would probably suffice. Then, features could be reduced using an iterative marginal information analysis approach to arrive at a set of informative loci and clinical features to be used. A predictive model trained on these features would function as a finished genetic test and could be validated and deployed.

This approach is described schematically in **Figure 3**.

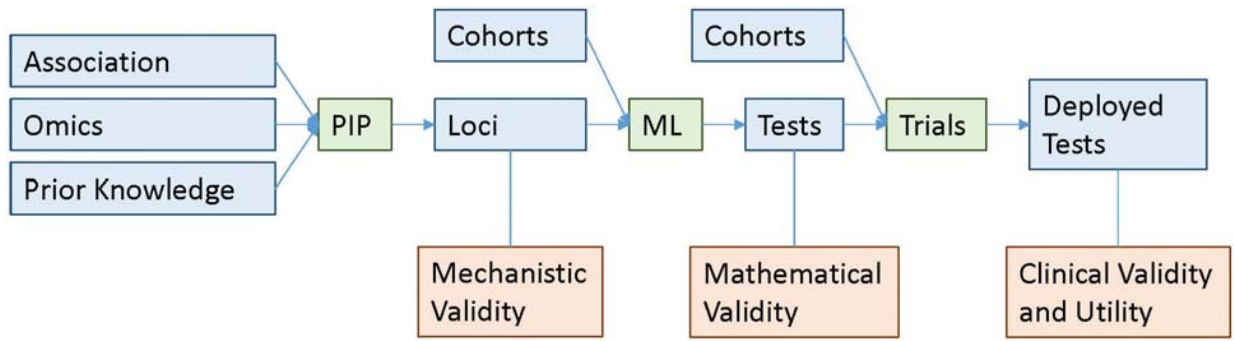

**Figure 3: Using PIP-Style Pipelines to Design Genetic Tests**
Schematic of an overall method for using PIP-style pipelines to design genetic-clinical tests. After loci are generated by a PIP experiment, they are used along with tractable and predictive clinical features in a cohort dataset to generate a predictive machine learning model, which is tuned with ablation analysis to find features with high marginal information. The minimal set of such features is used to generate a final predictive model which functions as a test, and can be validated in prospective and retrospective clinical trials.

Although historic regulatory approaches have demanded prospective clinical trials before clinical deployment of genetic tests and genetic prediction modules for clinical decision support, scientifically there is little conceptual distinction between a retrospective trial on a cohort which was not used in the construction of the test, and a prospective trial. As the scientific consensus around these issues congeals and makes its way into the culture of regulatory thinking, it is likely that the greater speed and lower expense of this approach, along with the clinical utility of making genetic testing more widespread, will carry the day in favor of this kind of development.

But in any event, even if prospective clinical trials were required in order to validate such predictive models, the fact that they were initiated on the basis of PIP-discovered variants with mechanistic validity should advantage such classifiers in the regulatory environment, compared with agnostic machine learning approaches based on whole genomes.

Much has been made of regulatory barriers and cultural hesitance to use genetic information in some medical specialties as explanatory elements for the slow progress of pharmacogenomics testing deployment in many clinical specialties. Historically, however, the single biggest factor preventing pharmacogenomics testing from reaching the clinic in any given case has not been regulatory or cultural barriers but the clinical utility of the underlying prediction. In instances in oncology and neuropsychiatry where such tests have added value, they have typically met with at least enough regulatory permission and cultural tolerance to be applied. It may be anticipated that if, in the fullness of time, tests for new phenotypes do add such utility, neither governments nor conservative clinicians will stand in their way indefinitely. And as the number of domains where such tests add value grows, they may become the object of much enthusiasm.

**Using PIP-style Pipelines to Construct a Phenome-wide Pharmacoepigenomic Atlas**

A PIP-style pipeline analysis, intended to discover variants with a causative role in a particular phenotype, takes place using a large amount of omics data which comes from the same database for each experiment, and is selected on the basis of preexisting knowledge about the phenotype under investigation. In the current version of the PIP, this principally comprises the MVF and MTF, containing information about the key mechanistic genes and associated variants, and the relevant tissues. In the case of the evolved featureset discussed in this perspective, other key information will come to the fore, including the most relevant Hi-C datasets, the degree of relatedness of clustered phenotypes, and the tissues for each phenotype. But regardless, a five box model pipeline requires a certain set of specific information about a phenotype system in order to run.

Many of the underlying elements of such a pipeline have begun to parallelize across all domains. For example, GWAS have now been conducted on thousands of phenotypes **[MacArthur et al 2017]**, and genotyped biobanks have enabled PheWAS **[Denny et al 2010]** methods to conduct parallel GWAS on thousands of phenotypes with one large cohort. Epigenome atlases now address an increasingly large cross section of the human body, with increasing granularity. Biomedical ontology databases now attempt to contain the knowledge on relatedness of phenotype categories and the contributions of tissues to phenotypes within a structured, computable vocabulary. And research in the design of EHR parsers is now proceeding **[Denny et al 2013, Beaulieu-Jones et al**

**2016]**, gesturing to a future world wherein individual health records can be grouped into phenotypic classes on a reliable and automated basis for an ever larger span of phenotypes.

The maturity of these trends will culminate in a world wherein, instead of performing a targeted PIP-style analysis on a particular phenotype of interest and using it to design a test, the converged datasets and methods described here will be used to perform parallel analyses and parallel test design for thousands of phenotypes. The predictors for these phenotypes could, for genotyped patients, then be contributed directly to an EHR system and used for clinical decision support. Furthermore, the genotypes, clinical records, and outcomes of patients within an EHR system could be used to refine the predictors for such a system on an ongoing basis.

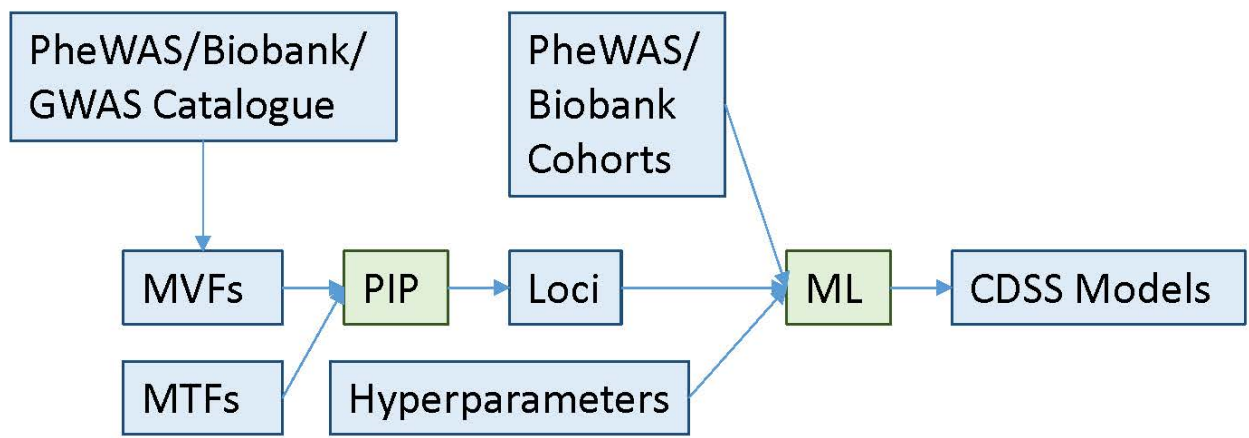

**Figure 4: Conceptual Process for Constructing a Pharmacophenomic Atlas**
Schematic representation of a conceptual process for parallelizing a PIP-style pipeline across thousands of phenotypes to create a genome-wide, phenome-wide pharmacophenome atlas. With either automation of parallelized manual work, MVFs and MTFs are created for thousands of phenotypes on the basis of PheWAS, biobanks, and/or the GWAS catalog, and thousands of PIP experiments run. Then, automated test generation as described in **Figure 3** is undertaken in parallel for all the phenotypes based on separate cohorts from a biobank and/or PheWAS. Finally, this set of thousands of predictive models for a comprehensive collection of pharmacological phenotypes may be used in every CDSS and research context wherein it may add value.

The concluding section of this perspective will describe a future vision for the means by which this process could take place, which is also pictured schematically in **Figure 4**.

The first step in such an analysis would be the construction of a phenotype set for investigation. Each PheWAS analysis already constructs such a set, typically manually.

The construction of tissue files on an automated basis requires a degree of natural language processing. Ontologic mapping of phenotypes to relevant diseases, along with crawling of structured literature databases like the EBI GWAS catalog **[Macarthur et al 2017]**, and natural language processing, could be used to create variant input files on an automated basis, by locating the relevant phenotypes in natural language and extracting the locus and population information. In addition to this, the output of a PheWAS on the same set of phenotypes could be used to construct the variant files.

Similarly, ontologies mapping drugs and diseases to their sites of action (e.g. SNOMED **[Millar 2016]**), cell lines to their cytologic properties and tissues of origin (e.g. Cell Line Ontology **[Sarntivijai et al 2014]**), organs and tissues in a hierarchical tree (e.g. NeuroFMA **[Turner et al 2010]**, HOMER **[Zhang et al 2011]**, BRENDA **[Gremse et al 2011]**), and synonymous tissue names to each other, may allow the creation of tissue files on an automated basis. This would create a description of all the relevant tissues for each phenotype.

With this done, it would be possible to run a PIP-style pipeline in parallel on many drug-disease systems to create a genome wide, phenome wide atlas of significant pharmacogenomic loci and gene networks. In particular, such experiments could be conducted in parallel on the basis of association data from EHR records for large populations **[Denny et al 2013]**. This atlas could include all the drug disease systems, disease risks, and other pharmacological phenotypes for which the underlying genetic associations and tissue specific omics are available.

Theoretically, if such automated methods as are described above proved impractical, and if sufficient resources were available, the manual curation of a library of tissue and variant files could be undertaken for a large library of phenotypes. The creation of input files for the current version of the PIP has already been reduced to a matter of days for two investigators, and with a suitable graphical interface for curation, could be reduced further, perhaps to the point where manual curation by a small dedicated staff with access to a variety of medical specialists became a tractable alternative. Certainly this has been the predominant approach in PheWAS design.

With a separate linked biobank not used in the original analysis, such a "PIP-WAS" atlas could be used to design predictive models for every phenotype under investigation. This would require the phenotypic extraction, key clinical variables, and clinical variable extraction for each phenotype from the second biobank to be automated as well. In addition, it would require the test generation and marginal information ablation analysis to be automated, a requirement which would require a lot of hyperparameter tuning. Nevertheless we are confident it is possible with current methods.

This comprehensive pharmacogenomics atlas would represent the industrialization of pharmacogenomics. It would enable, among other things, the scalable parallel design of tests for many diseases and drugs on the basis of genotype-linked EHR data, and the development of microarrays or sequencing panels containing the genetic information for many tests, or all, in one package. Such information could then be available in the EHR for a variety of purposes, and could even be present on, e.g., a military "dog tag" in the form of a QR code or other linked identifier for use in emergency and traumatic care settings.

Nor, indeed, would the construction of such an atlas need to be a discrete versioned event. If genotyped EHR data were added to the system, predictive models could refresh on an ongoing basis with new information, and new phenotypes could be added to the catalogue as their relevance was established.

**Pharmacogenomics in 2030**

What would medicine look like, in a world wherein such an atlas had been constructed, and wherein it had been incorporated into the EHR systems in use in the clinic, and in which patient genotypes were routinely available? Every time a patient required general health guidance, clinicians would have access to predictive information suggesting the diseases and syndromes which might present the greatest risk. For every prescription decision, the pharmacogenomics of the various drugs available for such an indication could be displayed, along with the particular adverse events most concerning for each one, for the individual patient. And for public health concerns with rare indications, the application of an optimized classifier to the covered population of a health system could prospectively identify a cohort who would benefit from prophylactic guidance or treatment.

Only twelve years ago, before the advent of GWAS, genetic tests were designed manually for each phenotype on the basis of painstaking, locus-specific mechanistic work, almost exclusively using coding variants, and then loci were assayed individually before being reported, often with a delay of weeks. While the clinical side of this picture has seen only muted change in the time since, rapid and fundamental advances on the research side, including omics atlases, PIP-style omics pipelines, spatial interaction data, biobanked EMRs, and PheWAS, have made it possible to see forward to a future, twelve years from now, when the parallel design of genetic tests for thousands of phenotypes, incorporating tissue-specific regulatory variants, machine learning, and large cohort datasets, has made the display of genetic predictive information on relevant phenotypes a routine and automatic part of life in the health care clinic.